# Evaluating Reputation Systems for Agent Mediated e-Commerce

Vibha Gaur, Neeraj Kumar Sharma, Punam Bedi
Department of Computer Science
University of Delhi, Delhi, India
3.vibha@gmail.com, neerajraj100@gmail.com, pbedi@du.ac.in

*Abstract*—Agent mediated e-commerce involves buying and selling on Internet through software agents. The success of an agent mediated e-commerce system lies in the underlying reputation management system which is used to improve the quality of services in e-market environment. A reputation system encourages the honest behaviour of seller agents and discourages the malicious behaviour of dishonest seller agents in the e-market where actual traders never meet each other. This paper evaluates various reputation systems for assigning reputation rating to software agents acting on behalf of buyers and sellers in e-market. These models are analysed on the basis of a number of features viz. reputation computation and their defence mechanisms against different attacks. To address the problems of traditional reputation systems which are relatively static in nature, this paper identifies characteristics of a dynamic reputation framework which ensures judicious use of information sharing for inter-agent cooperation and also associates the reputation of an agent with the value of a transaction so that the market approaches an equilibrium state and dishonest agents are weeded out of the market.

## I. INTRODUCTION

There is a vast amount of information available on WEB that is heterogeneous and distributed. This makes it practically infeasible for any user to combine all of the possible information sources to obtain an optimized and satisfactory solution to his problems. The software community addressed this challenge by developing intelligent software agents that act autonomously on behalf of users for jobs like shopping for the best priced and high quality products. Software agents offer a promise to change e-commerce trading by helping internet traders to purchase products from online distributed resources based on their interests and preferences [13]. Multi-agent systems (MAS) provide an environment populated with a set of self interested agents that cooperate to facilitate a complex process like e-commerce in a decentralized way. In the internet based e-market where actual traders never meet each other, reputation systems act as indispensable MAS to allocate reputation to other agents based on their past behaviour. Intelligent information sharing among agents in electronic markets is an important ingredient for either cooperating or competing in order to fulfil the purpose of maximizing their gains. Choosing ways to share information and to find the trustworthiness of other agents are the core activities of reputation systems.

Definitions of reputation vary with applications and contexts. From an objective point of view, reputation is expressed as "a quantity derived from the underlying social network which is globally visible to all members of the network" [2] or, "a perception that an agent has of another's intentions and norms" [5].

E-market environment in which these agents operate is generally open, that means agents can join or leave the marketplace at any time; uncertain, that is the true worth of a good can be judged only after purchase; and un-trusted, that is the e-market comprises of honest/dishonest agents. Designing efficient and secure reputation models that fulfil the requirement of both the parties involved in a transaction is a challenge in the research community. In literature different models are proposed that are based on centralized, distributed or hybrid architectures. Some models are based on direct evidence [1, 15] that means these assign reputation to a seller solely on the basis of their past interactions, while other models [3, 7, 12] also use shared opinion of other agents in MAS. These models employ Bayesian Probability Analysis [7], Reinforcement Learning [1, 15], Neural Networks [6] and Fuzzy Logic [12] for computation of reputation.

The objective of this paper is to evaluate various reputation systems from the literature based on their features and performance against different attacks. The systems described in literature are static in nature as these do not vary the relative weightage of individual and shared reputation components in the overall computation of a reputation value along with the change in experience of agents in the e-market. Most of these models do not take into account the size of a transaction while assigning a reputation value to a seller agent. To address these problems, this paper describes the characteristics of a framework for dynamic reputation system which is sensitive to the experience of agents involved in transactions and the value of a transaction in the e-market environment. This framework employs judicious use of information sharing and thus reduces the associated cost with effective inter-agent communication.

The rest of this paper is organized as follows. Various reputation models from literature are presented in section II and commercially available reputation systems are introduced in Section III. Section IV is used to evaluate these reputation systems based on their features and performance against different attacks. To address existing problems, important characteristics of a dynamic reputation framework are described in section V. Section VI comprises of a case study. Section VII concludes the paper with future research directions.

## II. REPUTATION MODELS

Reputation models are an important component of e-markets, help building trust and elicit cooperation among loosely connected and geographically dispersed economic agents [8]. A number of reputation models described in literature are discussed below in brief.

Evidential model [3, 4] is a reputation system that is based on a distributed reputation environment and Dampster Shafer Theory. Each agent is fully autonomous and has the capability to deal with fraud and deception by dishonest agents. An agent finds the trustworthiness of another agent [4] based on its direct interaction and testimonies given by other trustworthy agents.

Various reputation models [1, 15] described in the literature are based on reinforcement learning. In the reputation model for increasing user satisfaction [1], buyer agents assign the reputation to seller agents and, seller agents adjust the price and quality of products to maximise their profit. A multi-facet reputation model [15] involves reputation computation of both buyer and seller agents using quality, price and delivery time of goods. Both of the reputation models [1, 15] described above are based on direct evidence only.

REputation in GREgarious socieTies (REGRET) [12] employs fuzzy rules to classify the reliability of the witness agents based on their relationship with the target agent. REGRET is based on multi-facet reputation mechanism comprising three dimensions of reputation, namely individual, social and ontological. The reliability of a reputation value is modelled in this system based on the number of impressions/interactions of witness agents with the target agent.

Trust and Reputation model for Agent-based Virtual OrganizationS (TRAVOS) is a system based on Bayesian probability analysis [7]. Trust of an agent is modelled in this system by taking into account past experience between the two agents, and in case of lack of past experience, this model utilizes the information collected from third parties. To filter out unfair opinions, TRAVOS uses an endogenous approach by considering the statistical properties of the reported opinions alone.

A reputation model called "Truntis" [9] is based on accumulation of trust units (truntis). In this model, a seller must possess sufficient number of truntis before executing a transaction. To engage in a transaction, a seller agent must risk a particular quantity of truntis to cover the sale which is put into an escrow with the market operator. After a transaction, if a buyer is satisfied, seller gets more trunits, otherwise it loses risked truntis.

A flexible reputation and trust model [6] based on Artificial Neural Networks (ANN) employs the learning capability of backpropagation algorithm. This model tunes the parameters automatically to adapt to different personal requirements using ANN.

## III. COMMERCIAL REPUTATION SYSTEMS

A number of online reputation systems are in commercial use. eBay is the most popular auction site on the internet that has feedback forum as a reputation system in which after each transaction, a buyer gets an opportunity to rate a seller in the form of feedback as positive, negative or neutral i.e. +1, -1 or 0 respectively. All ratings received by an eBay user from other users are added up into a feedback rating number. The overall reputation of a user is computed by subtracting total number of negative feedbacks from the total number of positive feedbacks obtained from distinct users.

Amazon is America's largest online retailer. Initially, one becomes a member by first signing up. Reviews include star ratings from 1 to 5 and a prose text. Average of all ratings is used to assign reputation rating to a user.

## IV. EVALUATION OF REPUTATION SYSTEMS

The goal of a reputation system in an agent oriented e-commerce is to develop trustworthiness or the degree to which one agent/party has confidence in another within the context of a given purpose or decision [14]. A reputation system must ensure that after a number of transactions, the market reaches an equilibrium state and dishonest agents are weeded out of the market.

### A. Feature Analysis

This section evaluates the reputation systems based on characteristics including the range of assigned reputation values that enumerates the set of all possible reputation values of an agent in a model; reputation mechanism namely direct evidence/aggregation indicating that whether a model uses individual or aggregation of individual and shared opinion for computing the reputation value. These systems are further analysed on the basis of whether reputation is treated as a uni-facet or a multi-facet entity and whether recent ratings have more weightage than previous ratings. Table I summarizes results of analysis of various reputation systems described in sections II and III.

TABLE I. ANALYSIS OF REPUTATION SYSTEMS

| Reputation System | Reputation Range | Aggregation of Opinions | Computation Method | Other |
|---|---|---|---|---|
| eBay | +ve/neutral /-ve, -1/0/1 | Statistics and full data | Sum of all ratings | Rating based on feedback |
| Evidential Model | Numeric, [0,1] | Yes (Weighted Average) | Dempster-Shafer Theory | Referral chain bound by depth limit |
| Improving User Satisfaction | Numeric, [-1,1] | No (uses only direct experience) | Reinforcement learning | Adjustable product price & quality |
| E-market based on reputation | Numeric, [-1,1] | No (uses only direct experience) | Reinforcement learning | Multi-facet reputation calculation |
| TRAVOS | Probability, [0,1] | Yes (+ve, -ve experience) | Bayesian Probability Analysis | Estimate probability of accuracy |
| REGRET | Numeric, [-1,1] | Yes (weighted average) | Statistics, Fuzzy Inference | Multi-facet reputation, Ageing of rating |
| Broker-Assisting TRS based on ANN | Satisfy / Dissatisfy, 0/1 or [0,1] | Yes | ANN, Back-propagation, Clustering | Manages sub-communities, Ageing of ratings |
| Trunits Model | Numeric | No | Simple Math. Equations | Must possess truntis |

The reputation systems described above are relatively static in nature as in these systems, the weightage of individual and shared reputation is fixed and does not change with experience of agents. In addition, these do not take into account the value of a transaction. The performance of various reputation systems against different attacks is discussed in the following section.

*B. Performance Analysis based on Common Problems / Attacks and Proposed Solutions*

Performance analysis of a reputation system includes the ability of a reputation system to counter different attacks. The impact of attacks against reputation systems is much more than just the manipulation of reputation values as these results into money fraudulently lost and ruined business reputations [14]. Different classes of problems/attacks are classified as:

- *Ballot Stuffing (BS) and Badmouthing (BM)*: In BS, a group of agents collude to rate a particular agent with abnormally high ratings, whereas in BM an agent is rated abnormally low.
- *RECiprocity (REC) and RETaliation (RET)*: In REC two agents rate each other with abnormally high ratings, whereas in RET, both the agents rate each other with abnormally low ratings.
- *Re-ENtry (REN)*: A low rated agent exits and re-enter the market with a new identity defeating the purpose of reputation assignment.
- *Reputation-Lag (RL)*: It refers to the lag i.e. time gap, before cheating results in reduced reputation. In this period, an agent gets unlimited chances to cheat.
- *Value-IMbalance Problem (VIM)*: In this problem, reputation earned or lost during a transaction is not related to the value of a transaction. A malicious seller behaves honestly for small transactions to gain reputation and then cheats in large transactions.
- *Sudden-Exit (SE)*: An agent cheats for a large transaction and immediately exits from the market.
- *Multiple-Identity (MI)*: A seller agent is able to open multiple accounts and sell the products honestly through some and dishonestly through others.

These attacks/problems and their proposed solutions are summarised in Table II.

## V. DISCUSSION

Reputation systems are oriented to encourage trustworthy behaviour, increase user satisfaction and deter dishonest participants by providing means through which reputation could be computed and disseminated [13]. The e-market environment in which reputation systems operate is dynamic as it changes continuously in terms of agents freely entering and exiting the market and the varying experience of agents. Therefore, with each new transaction, the importance of an individual experience of a buyer-seller pair should increase as compared to the opinion shared by other agents. Moreover, the economic worth of being honest or dishonest in a transaction cannot be judged without taking into account the value of a transaction as honest behaviour in a large transaction is more important than in a small transaction.

The reputation systems described in literature show a relatively static behaviour as they do not take into account the increase in mutual experience of a buyer-seller pair with each successive transaction. Among the reputation systems studied, none except "Truntis" takes value of a transaction into account while allocating a reputation value to a target agent. To be robust and of high utility, a reputation system should be dynamic in nature so as to adapt to the changing environment and the experience of agents involved in a transaction.

*A. Characteristics of a Dynamic Reputation Framework*

A framework for Dynamic Reputation System (DRS) is based on two important characteristics of software agents. Firstly, to incorporate the importance of increased experience of a buyer-seller pair with each successive transaction, relative weight of individual reputation must increase and that of shared reputation must decrease. Secondly, DRS is based on the concept of Effective Reputation Value (ERV) which associates reputation of an agent with actual value of a transaction.

Effective Reputation Value (ERV) may be defined as "The reputation value of an agent achieved after incorporating the effect/importance of value of a transaction in the computation of reputation". In this framework, the overall reputation (R) depends on individual reputation (IR) and shared reputation (SR). The computation of R using DRS is described in (1).

$$R = \alpha * IR + (1 - \alpha) * SR \qquad (1)$$

TABLE II. PROBLEMS/ATTACKS AND PROPOSED SOLUTIONS

| Reputation System | Problems /Attacks | Proposed Solutions in the Model | Remarks |
|---|---|---|---|
| eBay | RL, VIM, MI, SE, REC | Entry fee is charged from seller to reduce REN, Limits feedback from unique users to tackle BS | Strong effect of REC, as about 98% of feedback is positive. |
| Evidential Model | RL, SE, VIM, RN, MI | Variation between prediction and observed trustworthiness is used to take care of BS/BM | Discounts rating by malicious witnesses |
| Improving User Satisfaction | BS, VIM, SE, REN, RL | Not effected by RL, BS/BM as the model is based on direct experience only | Uses only direct experience for reputation |
| E-market based on reputation | BS/BM, RL, VIM, REN, SE | Not effected by RL and BS/BM as the model is based on direct experience only | Uses only direct experience for reputation |
| TRAVOS | VIM, SE, REN, MI | Uses endogenous method to filter out unfair ratings | Uses opinions only in case of less confidence |
| REGRET | RL, VIM, SE, MI | Variation in ratings is used to judge reliability of reputation to reduce BS/BM | Computes reliability of reputation |
| Broker-Assisting TRS based on ANN | RL, SE, VIM, MI | Uses clustering algorithm and long term communications to protect from malicious clients | Based on backpropagation algorithm |
| Trunits Model | SE, MI | Provides solution to VIM, BS and partial solution to REN | Limits value of a transaction |

Where α is the experience gain factor and $0 \leq \alpha \leq 1$. Initially, the value of α is 0 before the first transaction between a buyer-seller pair and with each successive transaction, it increases by a small fraction up to 1. The actual rate at which the value of α should increase is to be decided by domain experts. After sufficiently large number of transactions, as value of α approaches 1, R would only depend on IR and the weightage of SR would effectively become zero. Assuming the incremental rate of α=0.01, the change in percentage of the relative weightage of IR and SR is shown in Fig.1.

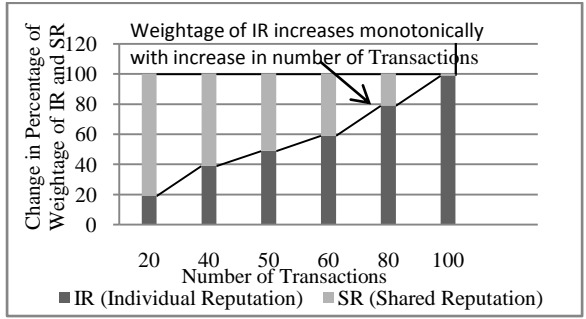

Figure 1. Change in relative weightage of IR and SR with increase in experience of a buyer-seller pair, assuming α incremental rate of 0.01

Once a good is purchased, the buyer agent either increases or decreases the reputation of a seller agent by incorporating the concept of ERV. Use of ERVs makes reputation of a seller agent monotonically proportional to the actual value of transaction by using η in (2) where η varies in the range [0,1] and is monotonically increasing with the value of a transaction.

$$\text{Change in Reputation} = \frac{\eta}{1+\beta} * R \quad (2)$$

$$\text{and,} \quad \eta = \frac{1}{1+e^{-\lambda x}} \quad (3)$$

Initially, β=0 before the first transaction between a buyer-seller pair and it is incremented by a small value with each successive transaction between a buyer-seller pair to a value specified by the domain expert, 'x' represents value of a transaction, λ represents a constant in the range [0,1]. The value of η represents reward/penalty for honest/dishonest behaviour of a seller agent during a transaction. By assuming current reputation i.e. R=0.45, λ=0.001 and β=0, change in reputation using (2) is shown in Table III and Fig. 2.

TABLE III. REPUTATION CHANGES MONOTONICALLY WITH THE CHANGE IN THE VALUE OF TRANSACTION

| Value of Transaction (x) | $\eta/(1 + \beta)$ (Assuming β = 0) | Change in Reputation |
|---|---|---|
| 100 | 0.524978 | 0.23624 |
| 400 | 0.598685 | 0.269408 |
| 800 | 0.689969 | 0.310486 |
| 1500 | 0.817568 | 0.367905 |
| 2500 | 0.924137 | 0.415861 |

Initially, when β is 0, $\frac{\eta}{1+\beta} = \frac{\eta}{1+0} = \eta$ and as the value of β increases with each successive transaction between a buyer-seller pair, the relative increase/decrease in reputation is discounted due to the convention that reputation gained from different buyers is more important than reputation gained due to successive transactions between the same buyer-seller pair. By assuming β=0.6 and keeping other parameters same as in Table III, reduced change in reputation as per (2) is shown in Table IV.

TABLE IV. REDUCED CHANGE IN REPUTATION WITH SUCCESSIVE TRANSACTIONS BETWEEN A BUYER-SELLER PAIR

| Value of Transaction (x) | $\eta/(1 + \beta)$ (Assuming β = 0.6) | Change in Reputation |
|---|---|---|
| 100 | 0.328112 | 0.14765 |
| 400 | 0.374178 | 0.16838 |
| 800 | 0.431231 | 0.194054 |
| 1500 | 0.51098 | 0.229941 |
| 2500 | 0.577585 | 0.259913 |

For β=0 and β=0.6 with incremental rate of β as 0.001, relative change in reputation is shown below in Fig. 2.

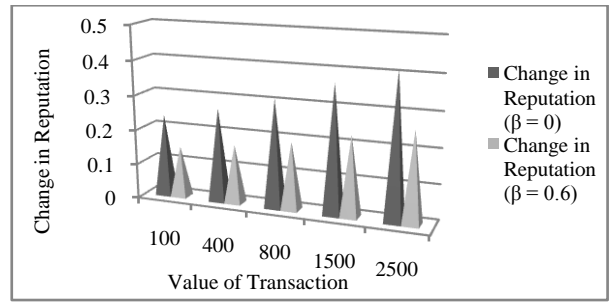

Figure 2. Relative change in reputation due to increase in value of transaction and number of transactions between a buyer-seller pair

This framework employs judicious use of information sharing and thus reduces the associated cost with effective inter-agent communication.

The success of a reputation system depends on its inbuilt defence mechanisms to counter different attacks. In DRS, the problem of Ballot Stuffing (BS) gets reduced with each successive transaction between a buyer-seller pair as the weightage of shared reputation decreases with increase in individual experience. Further, the effect of Reciprocation (REC) and Retaliation (RET) is also minimised in DRS as the change in reputation of seller due to transactions between the same buyer-seller pair is discounted with each successive transaction. DRS also reduces the Re-entry (REN) problem as to re-enter in the market as a new agent, an agent has to lose its current reputation and should start with minimum reputation. Further, DRS resolves Value Imbalance (VIM) problem, as the reputation earned is monotonically related to the value of the transaction.

## VI. CASE STUDY

A simple case study was conducted involving four buyers ($b_1$ to $b_4$) and six sellers ($s_1$ to $s_6$) to visualise two scenarios in e-market to verify the effect of participants' dishonest behaviour in a reputation system that is based on the characteristics of DRS.

Different sellers involved in a transaction face the moral hazard of behaving either honestly or dishonestly. Dishonest sellers may launch an attack on the reputation system to maximise their gains.

In a particular scenario, a buyer $b_1$ and seller $s_4$ were involved in repeated transactions and after 25 transactions, the reputation of seller $s_4$ is found to be 0.35. Further, buyer $b_1$ and seller $s_4$ interacted in six more transactions with seller $s_4$ behaving honestly in first five of value 700 each and cheated in the sixth transaction of worth 2000. Without the concept of ERV, amount of increase and drop in reputation due to the result of a transaction would have been same that would support dishonest agents that launch VIM attack. Due to the effect of ERV, the drop in the reputation rating of $s_4$ as a result of cheating is relatively greater due to the larger value of transaction that helped in resolving the VIM problem as shown in Fig. 3.

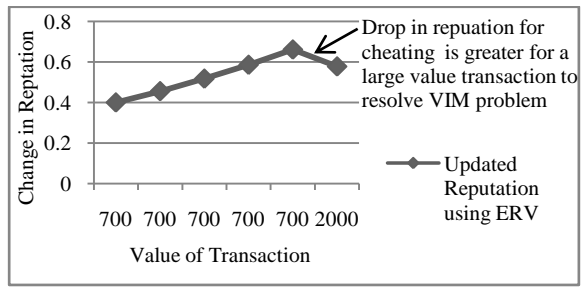

Figure 3. Drop in reputation is greater for a large value transaction to counter Value Imbalance Problem (VIM)

In another scenario, buyer $b_3$ is involved in transactions with seller $s_5$ with an existing reputation rating of 0.18. At this time Ballot Stuffing (BS) attack is launched on buyer $b_3$ due to which seller $s_5$ is given a high shared reputation (SR) of 0.98. The effect of BS is reduced with successive transactions between buyer $b_3$ and seller $s_5$ as the weightage of SR is decreased with each subsequent transaction seller with $s_5$. With the incremental value of α as 0.01 and β as 0.1 respectively, the effect of BS is reduced to zero after 100 transactions as shown in Fig. 4.

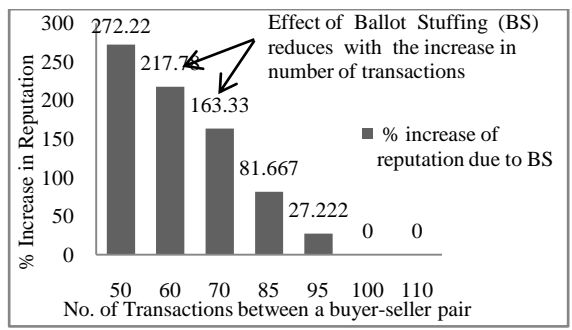

Figure 4. Reduced effect of Ballot Stuffing due to reduced weightage of SR with successive transactions between a buyer-seller pair

Similarly, the effect of Badmouthing (BM) would also be reduced with increase in number of transactions between a buyer-seller pair due to reduced weightage of SR in the overall reputation computation.

VII. CONCLUSION AND FUTURE WORK

Various reputation systems from literature were analysed in this paper with respect to different features viz. computational models and common attacks/problems on reputation systems along with their defence mechanisms. To address problems in these systems due to their relatively static nature, a general framework for dynamic reputation computation is proposed, which is sensitive to the changing parameters of e-market environment like experience of agents and value of a transaction in e-market environment. In this framework, increase in transactional experience leads to increased weightage of individual reputation and honesty in a large transaction leads to a greater increase in reputation as compared to a small transaction. Future directions to this work include developing algorithms for computing reputation of buyer/seller agents and enhancing this framework into a robust reputation system with improved capability to counter attacks on reputation systems.